\documentclass[twocolumn,showpacs,preprintnumbers,amsmath,amssymb]{revtex4}
\usepackage{graphicx}
\usepackage{dcolumn}
\usepackage{bm}
\def\bea {\begin{eqnarray}}
\def\eea {\end{eqnarray}}
\def\ra {\rightarrow}
\def\be {\begin{equation}}
\def\ee {\end{equation}}
\def\nn {\nonumber}
\begin{document}
\vskip 0.2in
\vskip 0.2in
\begin{flushleft}
{\bf Physical Review C (in press)}
\end{flushleft}

\title{Thermal photon to dilepton ratio in high energy nuclear collisions}

\author{Jajati K. Nayak, Jane Alam, Sourav Sarkar and Bikash Sinha}

\medskip

\affiliation{Variable Energy Cyclotron Centre, 1/AF, Bidhan Nagar , 
Kolkata - 700064}

\date{\today}

\begin{abstract}

The ratio of transverse momentum distribution of thermal photons to dilepton 
has been evaluated. It is  observed that this ratio reaches a plateau  beyond
a certain value of transverse momentum. 
We argue that this ratio can be used to estimate
the initial temperature of the system by 
selecting the transverse momentum and invariance mass windows
judiciously.  It is demonstrated that if the radial flow is large then
the plateau disappear and hence a deviation from the plateau 
can be used as an indicator of large radial flow.
The sensitivity of the results on various input parameters has been studied. 

\end{abstract}

\maketitle

\section{Introduction}
Collisions between nuclei at ultra-relativistic energies
produce charged particles - either in the hadronic or in the
partonic state, depending on the collision energy.
Interaction of these charged particles produce real and
virtual photons (lepton pairs). 
Because of their nature of interaction, the
mean free path of electromagnetic (EM) radiation 
is large compared to the size of the system
formed after the collision. 
Therefore, EM radiation 
can be used as an efficient tool to understand the
initial conditions of the system ~\cite{mclerran,gale,weldon,alam1,alam2}
and hence  can be used to probe 
the quark gluon plasma (QGP) formation in
heavy ion collisions (HIC). Practically, however, this is a 
difficult task, because on the  one hand the thermal radiation from QGP
has to be disentangled from  those produced in
initial hard collisions and from the decays of hadrons and
on the other hand the evaluation of thermal photon
and dilepton spectra need various inputs {\it such as}  
initial temperature ($T_i$), thermalization time ($\tau_i$), 
equation of state (EOS), 
transition temperature ($T_c$), freeze-out
temperature ($T_f$) etc, which are not known unambiguously.
The sensitivity of the photon spectra on these inputs are 
demonstrated in~\cite{anrms,assb}.
Therefore, theoretical results on transverse momentum ($p_T$)
spectra  of photons and dileptons
always suffer from these uncertainties. 
Of course, certain constraints can be imposed  on these inputs
from experimental results - {\it e.g.} transverse mass
spectra of hadrons  and hadronic multiplicities 
are useful quantities for constraining 
freeze-out conditions and  initial entropy production.

In the present work,
therefore, we evaluate the ratio of the transverse momentum spectra 
of thermal photons and lepton pairs: 
\be
R_{em} = (d^2N_\gamma/d^2p_Tdy)_{y=0}/(d^2N_{\gamma^\ast}/d^2p_Tdy)_{y=0} 
\label{eq1}
\ee
in which most of the 
uncertainties mentioned above are expected to get canceled so that
it provides accurate information~\cite{sinha,raha}
about the state of the matter formed initially.
We calculate the ratio, $R_{em}$ for SPS, RHIC 
and LHC energies.

The paper is organized as follows. 
In section II we discuss the invariant yield of thermal 
photons and lepton pairs . In section III, the space time 
evolution is outlined. Results are presented in 
section IV. Finally in section V, we summarize the work.
\section{Production of thermal photons and $e^+e^-$ pairs}
The rate of thermal dilepton production per unit space-time volume
per unit four momentum volume is given by\cite{mclerran,gale,weldon,alam2}
\be
\frac{dN}{d^4pd^4x}=\frac{\alpha}{12\pi^4M^2}L(M^2)
\mathrm{Im\Pi_\mu^{R\mu}}f_{BE}
\label{virtual}
\ee
$\alpha$ is EM
coupling, $\mathrm{Im\Pi_\mu^\mu}$ is the imaginary part of the 
retarded photon self energy and $f_{BE}$ is the Bose-Einstein
factor which is a function of $u^\mu p_\mu$ for a thermal system
having four velocity $u^\mu$ at each space-time point of 
the system,
$p^2(=p_\mu p^\mu)=M^2$ is the  invariant mass square of the lepton pair and
\be
L(M^2)=\left(1+\frac{2m^2}{M^2}\right)\sqrt{1-4\frac{m^2}{M^2}}
\label{spinor}
\ee
arises from the final state leptonic current involving Dirac
spinors. $m$ in Eq.~\ref{spinor} is the lepton mass.

The real photon production rate can be obtained from the
dilepton emission rate by replacing
the product of the EM vertex $\gamma^*\rightarrow l^+l^-$, the
term involving final state leptonic current and the square of the (virtual) 
photon propagator by the polarization sum 
($\sum_\mathrm{polarization}\epsilon^\mu\epsilon^\nu=-g^{\mu\nu}$) for the real
photon. Finally the phase space factor for the lepton-pairs should
be replaced by that of the photon to obtain the photon emission rate,
\be
E\frac{dN}{d^4xd^3p}=\frac{g^{\mu\nu}}{(2\pi)^3}\mathrm{Im\Pi_{\mu\nu}}f_{BE}
\label{real}
\ee
(see~\cite{mclerran,gale,weldon,alam2} for details).

The results given above is correct up to order $e^2(\sim \alpha)$
in EM interaction but exact, in principle, to all order in strong interaction. 
Now it is clear form Eqs.~\ref{virtual} and ~\ref{real} that
for the evaluation of photon and dilepton production rates one needs to
evaluate the imaginary part of the photon self energy.
The Cutkosky rules or thermal cutting rules give a systematic procedure
to express the imaginary part of the photon self energy in terms of the
physical amplitude.

\subsection{Thermal photons}
Ideally, one wants to detect photons from QGP. However, 
the experimental measurements contain 
photons from various processes e.g, from the hard collisions 
of initial state partons of the colliding nuclei, thermal 
photons from quark matter and hadronic matter and photons from 
the  hadronic decays after freeze-out. The contributions from the initial 
hard collisions of partons is under control via perturbative QCD (pQCD).
The data from $pp$ collisions will be very useful to validate pQCD calculations.
Photons from the hadronic decays ($\pi^{0}$ $\rightarrow$ $\gamma$$\gamma$, 
$\eta$ $\rightarrow$ $\gamma$$\gamma$ etc.) 
can be reconstructed, in principle,  by invariant mass analysis.
But the most challenging task is to separate the 
thermal photons originating from the hadronic phase,  
which needs careful theoretical estimation. 
\par
The invariant yield of thermal photons can be written as 
\begin{equation}
\frac{d^2N_\gamma}{d^2p_{T}dy}=\sum_{i=Q,M,H}{\int_{i}{\left(\frac{d^2R_\gamma}
{d^2p_{T}dy}\right)_id^4x}}
\label{eq2}
\end{equation}
where $i\equiv Q, M, H$ represents QGP, mixed (coexisting 
phase of QGP and hadrons) 
and hadronic phases respectively. 
$(d^2R/d^2p_{T}dy)_i$ is the static rate of photon 
production from the phase $i$, which is convoluted over 
the expansion dynamics
through the integration over $d^4x$. 
\subsubsection {Thermal photons from Quark Gluon Plasma}
The contribution from QGP 
to the spectrum of thermal photons 
due to annihilation ($q$$\bar{q}$$\rightarrow$$g$$\gamma$) and 
Compton ($q(\bar{q})g\rightarrow q(\bar{q})\gamma$) 
processes has been calculated in~\cite{kapusta,bair} using 
hard thermal loop (HTL) approximation~\cite{braaten}. Later, it was 
shown that photons from the processes~\cite{aurenche1}: 
$g$$q$$\rightarrow$$g$$q$$\gamma$, 
$q$$q$$\rightarrow$$q$$q$$\gamma$, 
$q$$q$$\bar{q}$$\rightarrow$$q$$\gamma$ 
and $g$$q$$\bar{q}$$\rightarrow$$g$$\gamma$ 
contribute in the same order $O(\alpha\alpha_s)$ as
Compton and annihilation processes.
The complete calculation of emission rate from QGP to order $\alpha_s$
has been performed by resuming ladder diagrams in the effective 
theory~\cite{arnold}. In the present work this rate has been used.
The temperature dependence of the strong coupling, $\alpha_s$ 
has been taken from~\cite{zantow}.

\subsubsection{Thermal photons from hadrons} 

For the photon spectra  from hadronic phase
we consider an exhaustive set of hadronic reactions and the radiative
decay of higher resonance states
~\cite{we1,we2,we3}.
The relevant reactions and decays for photon production are:
(i) $\pi\,\pi\,\ra\,\rho\,\gamma$, (ii) $\pi\,\rho\,
\ra\,\pi\gamma$ (with all possible mesons in the
intermediate state~\cite{we3}), (iii)$\pi\,\pi\,\ra\,\eta\,\gamma$ and
(iv) $\pi\,\eta\,\ra\,\pi\,\gamma$,
$\rho\,\ra\,\pi\,\pi\,\gamma$ and $\omega\,\ra\,\pi\gamma$.
The corresponding vertices's are obtained
from various phenomenological Lagrangians described in detail
in Ref.~\cite{we1,we2,we3}.
The reactions involving strange mesons:
$\pi\,K^\ast\ra K\,\gamma$,
$\pi\,K\ra K^\ast\,\gamma$,
$\rho\,K\ra K\,\gamma$ and
$K\,K^\ast\ra \pi\,\gamma$~\cite{turbide} have also
been incorporated in the present work.
Contributions from other decays, such as
$K^{\ast}(892)\,\ra\, K\,\gamma$, $\phi\,\ra\,\eta\,\gamma$,
$b_1(1235)\,\ra\,\pi\,\gamma$, $a_2(1320)\,\ra\,\pi\,\gamma$
and $K_1(1270)\,\ra\,\pi\,\gamma$ have been found to be
small~\cite{haglin} for $p_T>1$ GeV.
All the isospin
combinations for the above reactions and decays have properly been
taken into account. The effects of hadronic form factors~\cite{turbide} 
have also been incorporated in the present calculation. 
\subsection{Thermal dileptons}
Like photons, dileptons can also be used as an efficient probe
for QGP diagnostics, provided one can subtract out contributions
from Drell-Yan process, decays of vector mesons within the
life time of the fire ball and hadronic decays occurring 
after the freeze-out. Like hard photons, lepton pairs from 
Drell-Yan processes can be estimated  by pQCD.  
The $p_T$ spectra of thermal lepton pair suffer from the 
problem of indistinguishability between QGP and hadronic sources 
unlike the usual invariant mass ($M$) spectra which shows characteristic 
resonance peaks in the low $M$ region. The invariant 
transverse momentum distribution of thermal dileptons ($e^+e^-$ or
virtual photons, $\gamma^\ast$) is given by:
\begin{equation}
\frac{d^2N_{\gamma^\ast}}{d^2p_{T}dy}=\sum_{i=Q,M,H}{\int_{i}
{\left(\frac{d^2R_{\gamma^\ast}}{d^2p_{T}dydM^2}\right)_idM^2d^4x.}}
\label{eq3}
\end{equation}
The limits for integration over $M$ can be fixed 
judiciously to detect contributions from either 
quark matter or hadronic matter (see Fig.\ref{fig1}).
Experimental measurements~\cite{na60,phenix} are
available for different $M$ window.

\subsubsection{Dileptons from QGP}
In the plasma phase the 
lowest order process  producing lepton pair is 
$q \bar{q} \rightarrow \gamma ^{*}\rightarrow l^{+}l^{-}$. 
QCD corrections to this rate have 
been obtained for a QCD plasma at finite temperature in 
Refs.~\cite{altherr,thoma} up to  order $O(\alpha^2 \alpha_s$). 
In the present work contribution up to $O(\alpha^2\alpha_s)$
has been considered.

\subsubsection{Dileptons from Hadrons} 
The following parametrization~\cite{alam2,shuryak} has been used to evaluate
the  dilepton emission rates from light vector mesons ($\rho$, $\omega$
and $\phi$):
\begin{eqnarray}
\frac{d^2R_{\gamma^\ast}}{dM^2d^2p_Tdy}&=&\frac{\alpha^2}{2\pi^3}f_{BE}
[\frac{f_{V}^2M\Gamma_{V}}{(M^2-m_{V}^2)^2+ (M\Gamma_{V})^2}\nn\\
&&+\frac{1}{8\pi}\frac{1}{1+exp((w_0-M)/\delta)}\nn\\ 
&&\times (1+\frac{\alpha_s}{\pi})].
\label{eq4}
\end{eqnarray}
These parameterizations are consistent with the 
experimental data from $e^+\,e^-\,\ra\,V (\rho,\omega$ or $\phi)$ 
processes~\cite{shuryak,alam2,rw}.
Here, $f_{BE}$, is the Bose-Einstein distribution. 
$f_V$ is the coupling  between the EM current and
vector meson fields, $m_V$ and $\Gamma_{V}$ are the masses and widths of 
the vector mesons and $\omega_0$ is the continuum threshold 
above which the asymptotic freedom is restored.  
We have taken 
$\alpha_s=0.3$, $\delta=0.2 GeV$, $\omega_0=1.3$  GeV for $\rho$ 
and $\omega$. For $\phi$ we have taken $\omega_0=1.5$ GeV and 
$\delta=1.5$ GeV. The EM current in terms of $\rho$,
$\omega$ and $\phi$ field can be expressed as $J_\mu=J_\mu^\rho
+J_\mu^\omega/3- J_\mu^\phi/3$. Therefore, the contributions from
$\omega$ and $\phi$ will be down by a factor of 9.

\section{Space-time evolution}
The matter formed after ultra-relativistic heavy ion
collisions undergo space-time evolution, which can
be described by relativistic hydrodynamics. In the
present work
the space time evolution of the system 
has been studied using ideal relativistic hydrodynamics 
in (2+1) dimension ~\cite{von} with longitudinal boost 
invariance ~\cite{bjorken} and cylindrical symmetry. 
The initial temperature($T_i$) 
and thermalization time ($\tau_i$) are constrained 
by the following equation ~\cite{Hwa} for an isentropic expansion:
\be
T_i^{3}\tau_i \approx \frac{2\pi^4}{45\xi(3)}\frac{1}{4a_{eff}}\frac{1}{\pi R_A^2}\frac{dN}{dy}.
\label{eq6}
\ee
where, $dN/dy$= hadron multiplicity,  
$R_A$ is the radius of the system, 
$\xi(3)$ is the Riemann zeta function and  $a=\pi^2g/90$
($g=2\times 8+ 7\times 2\times 2\times 3\times N_F/8$) 
is the degeneracy of the massless quarks and gluons in the QGP, 
$N_F$=number of flavours.
The values of initial temperatures and thermalization times
for various beam energies are shown in table I.
The initial energy density,$\epsilon(\tau_i,r$) and radial velocity,
$v_r(\tau_i,r)$  profiles are taken as:
\be
\epsilon(\tau_i,r)=\frac{\epsilon_0}{1+e^{\frac{r-R_A}{\delta}}} 
\label{eq7}
\ee
and
\be
v_r(\tau_i,r)=v_0\left(1-\frac{1}{1+e^{\frac{r-R_A}{\delta}}}\right),  
\label{eq8}
\ee

where the surface thickness, $\delta=0.5$ fm. 
We have  taken
$v_0=0$,  which can reproduce the measured hadronic spectra 
at SPS and RHIC energies~\cite{npa2002,anrms}. So far there
is no consensus on the value of 
$T_c$, it varies from 151 MeV~\cite{fodor} to 192 MeV~\cite{cheng}.
In the present work we assume $T_c = 192$ MeV. 
In a first order phase transition scenario - 
we use the bag model EOS for the QGP phase and for the hadronic 
phase all the resonances with mass $\leq 2.5$ GeV have been 
considered~\cite{bm}.  

To show the sensitivity of the results on the EOS 
we also use the lattice QCD EOS for $T\geq T_c$~\cite{MILC}. 
For  the hadronic matter (below $T_c$)
all the resonances with mass $\leq 2.5$ GeV have been considered
~\cite{bm}. For the transition region the following parametrization
has been used~\cite{hatsuda}.
\be
s=f(T)s_q + (1-f(T))s_h
\ee
where $s_q$ ($s_h$) is the entropy density of the quark (hadronic) 
phase at $T_c$ and 
\be
f(T)=\frac{1}{2}(1+tanh(\frac{T-T_c}{\Gamma}))
\label{eos}
\ee
the value of the parameter $\Gamma$ can be varied to
make the transition strong or weak first order. Results for various
values of $\Gamma$ are given below. 

\begin{table}
\caption{The values of various parameters - thermalization
time ($\tau_i$), initial temperature ($T_i$), freeze-out temperature
($T_f$) and hadronic multiplicity $dN/dy$  - used 
in the present calculations.}
\begin{tabular}{lcccr}
\tableline
Accelerator &$\frac{dN}{dy}$&$\tau_i(fm)$&$T_i$(GeV) &$T_f$ (MeV)\\
\tableline
SPS&700&1&0.2&120\\
RHIC&1100&0.2&0.4&120\\
LHC&2100&0.08&0.7&120\\
\tableline
\end{tabular}
\end{table}
\begin{figure}
\begin{center}
\includegraphics[scale=0.45]{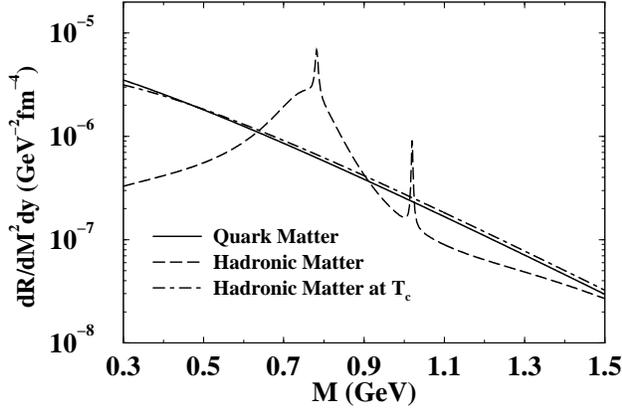}
\caption{The invariant mass distributions of thermal dileptons
from QGP and hadronic matter at $T=200$ MeV. Solid (dashed) line
indicates the emission rates from QGP (hadronic matter).
The dot-dashed line stands for emission rate from hadronic matter
at the transition temperature (see text).
} 
\label{fig1}
\end{center}
\end{figure} 

\begin{figure}
\begin{center}
\includegraphics[scale=0.45]{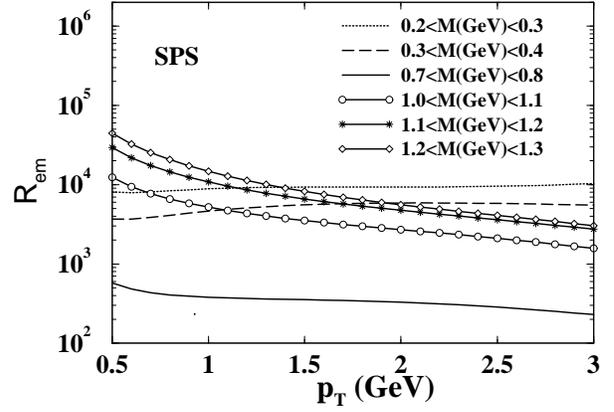}
\caption{The thermal photon to dilepton ratio, $R_{em}$  as  a function 
of transverse momentum, $p_T$ for various invariant mass window. 
} 
\label{fig2}
\end{center}
\end{figure} 
\begin{figure}
\begin{center}
\includegraphics[scale=0.45]{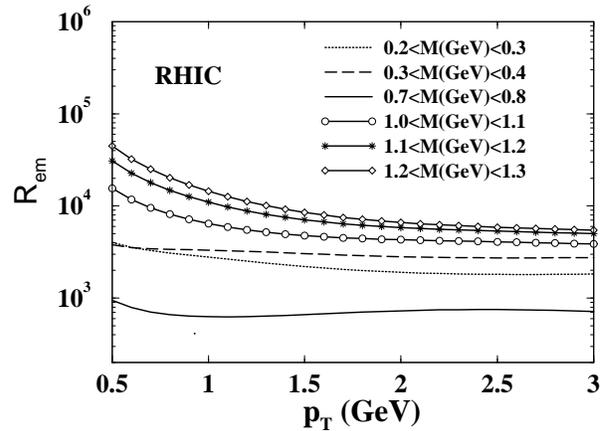}
\caption{Same as Fig.~\ref{fig1} for RHIC energy}
\label{fig3}
\end{center}
\end{figure} 
\begin{figure}
\begin{center}
\includegraphics[scale=0.45]{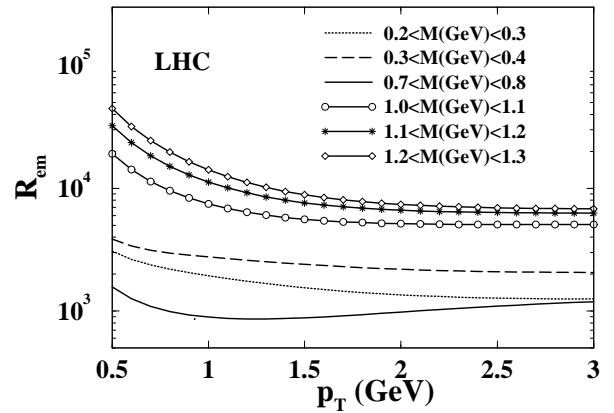}
\caption{Same as Fig.~\ref{fig1} for LHC energy}
\label{fig4}
\end{center}
\end{figure} 
\begin{figure}
\begin{center}
\includegraphics[scale=0.45]{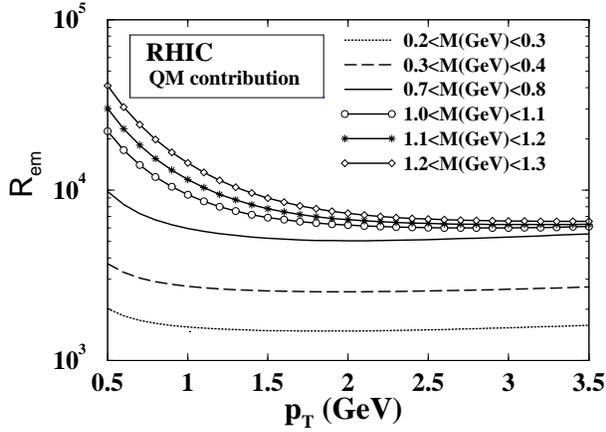}
\caption{Same as Fig.~\ref{fig2} for quark matter phase only.
}
\label{fig5}
\end{center}
\end{figure} 
\begin{figure}
\begin{center}
\includegraphics[scale=0.45]{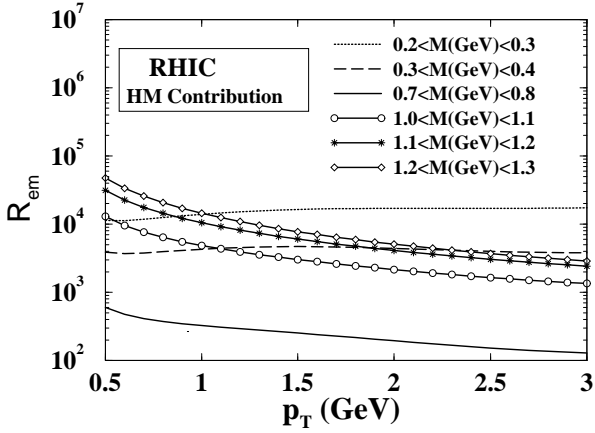}
\caption{Same as Fig.~\ref{fig2} for hadronic phase only.
}
\label{fig6}
\end{center}
\end{figure} 
\begin{figure}
\begin{center}
\includegraphics[scale=0.45]{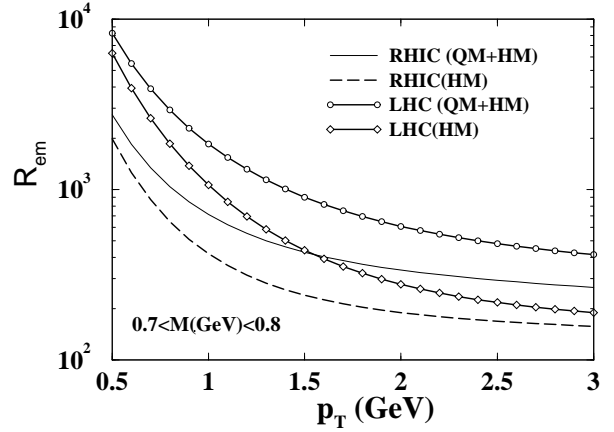}
\caption{The variation $R_{em}$ with $p_T$  
for invariant mass window, $M=0.7-0.8$ GeV. An unrealistically
large value to radial flow has been given initially to demonstrate that 
large flow can destroy the plateau structure of $R_{em}$. Other inputs are similar to those of Figs.\protect\ref{fig3} and \protect\ref{fig4}. 
}
\label{figflow}
\end{center}
\end{figure} 
\begin{figure}
\begin{center}
\includegraphics[scale=0.45]{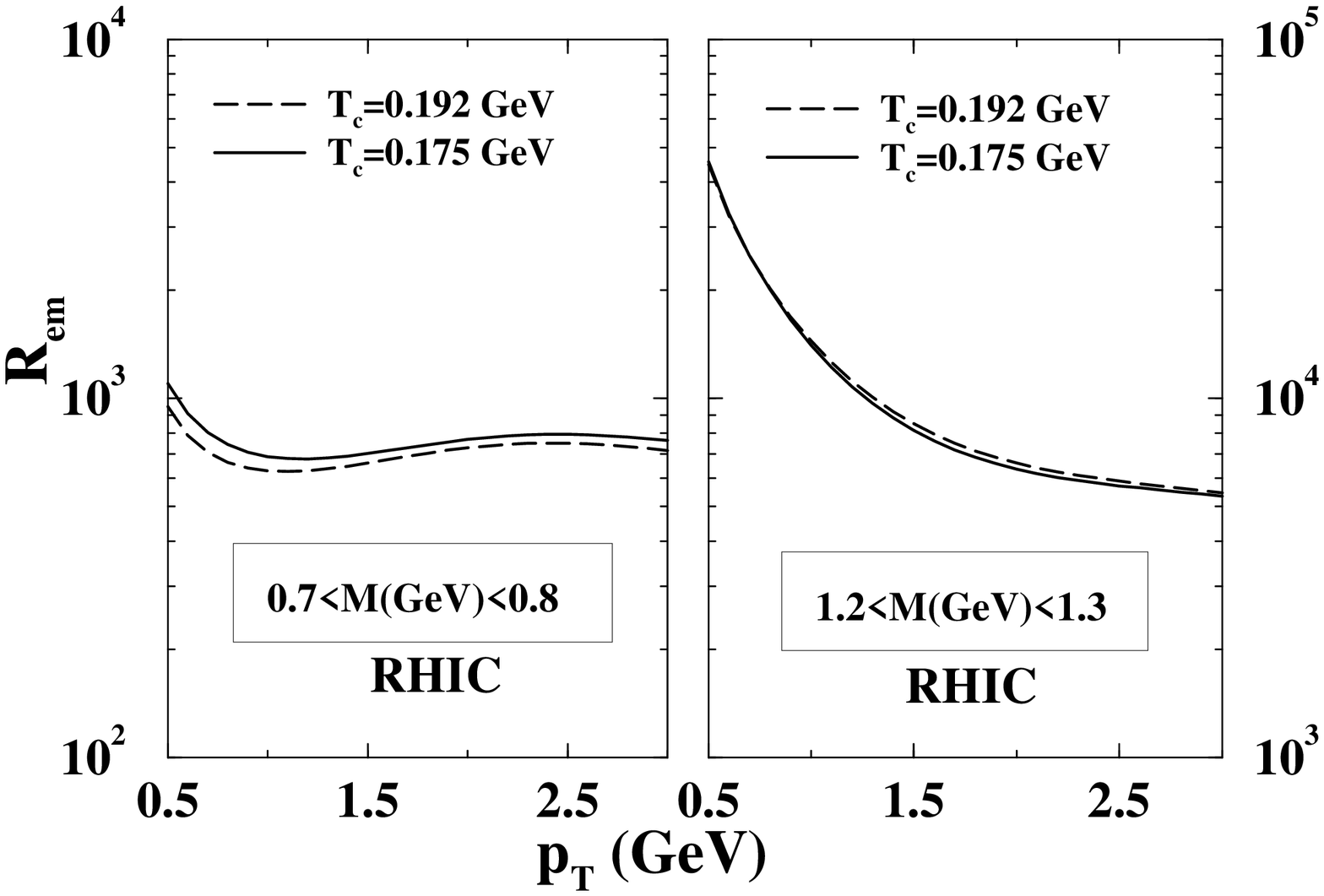}
\caption{$R_{em}$ as a function of $p_T$ for different values of $T_c$ 
for invariant mass windows, $M=0.7-0.8$ GeV and 
$M=1.2-1.3$ GeV.
}
\label{fig7}
\end{center}
\end{figure} 
\begin{figure}
\begin{center}
\includegraphics[scale=0.4]{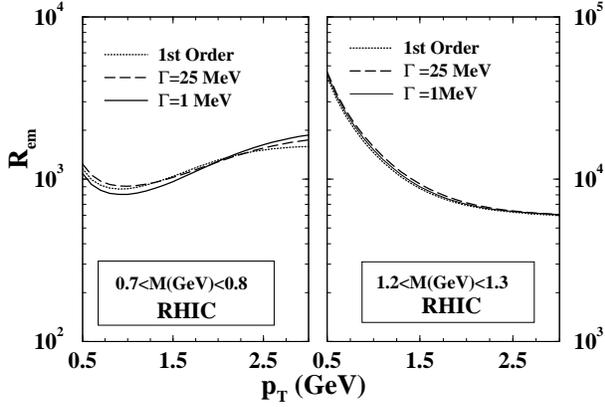}
\caption{$R_{em}$ as a function of $p_T$ for different EOS for 
invariant mass windows, $M=0.7-0.8$ GeV  and
$M=1.2-1.3$ GeV.
}
\label{fig8}
\end{center}
\end{figure} 
\begin{figure}
\begin{center}
\includegraphics[scale=0.45]{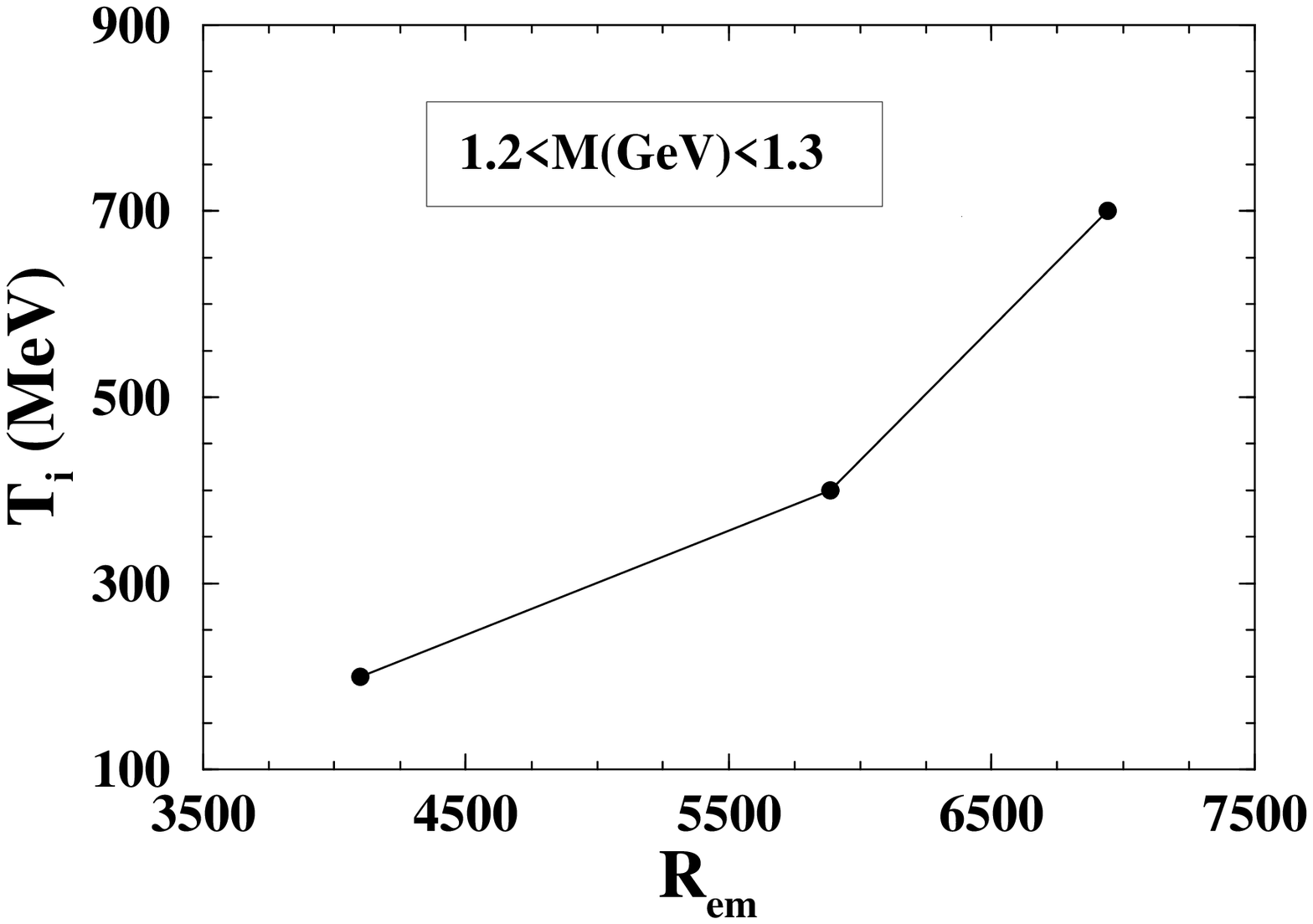}
\caption{Initial temperature is plotted as function 
$R_{em}(p_T=2.5\mathrm{GeV})$ for
the  $M$  window, $M=1.2-1.3$ GeV. 
}
\label{figti}
\end{center}
\end{figure} 
\section{Results}
The values of the initial and freeze-out parameters 
shown in table I  along
with the EOS mentioned above  have been used 
as inputs to hydrodynamic calculations. 
The experimental
data from SPS on hadrons~\cite{na49},  photons~\cite{wa98} and 
$M$ distribution of dileptons~\cite{ceres} have been reproduced    
in~\cite{npa2002}, ~\cite{ja1,ja2} and~\cite{sah} respectively 
by using these inputs. 
The values of the initial parameters for SPS agree 
with the results obtained from the analysis of photon spectra in 
Refs.~\cite{steffen,huovinen,KG,DYP}.
Recently the data from PHENIX collaboration at RHIC~\cite{adler,buesching}
has also been explained in~\cite{anrms} (see also~\cite{peress}) 
with the parameters mentioned in  table I. 
The lepton pairs measured
by NA60 collaboration~\cite{na60prl,na60prl2} in In-In collisions 
has been explained by spectral broadening of
$\rho$~\cite{hees,ruppert}. 

The emission rate from hadronic and quark matter at a temperature
of  200 MeV has been displayed in Fig.~\ref{fig1}. The contribution
from QGP dominates over its hadronic counterpart (without 
any medium effects) for $M<600$ MeV
and  $M>1.1$ GeV, therefore, these windows are better suited
for the detection of QGP.
However, it should be mentioned here that the modification 
of the spectral functions of vector mesons (especially $\rho$
and $\omega$) - pole shift~\cite{BR} or broadening~\cite{rw}
may give rise to dileptons at the lower $M$ region making it difficult
to detect contributions from QGP.
The change in the hadronic spectral function  will enhance
the dileptons from the hadronic contribution in the lower
mass ($M<600 MeV$) window, however the overall structure in
the ratio, $R_{em}$ will not change appreciably.
At the transition temperature ($\sim 200$
MeV) if one assumes the vector mesons
masses go zero {\it a la} Brown-Rho scaling~\cite{BR} then 
all the peaks in the dilepton 
spectra disappeared and the rates obtained from EM 
current-current correlator (dot-dashed line) 
are close to the rate from QGP, indicating that
the $q\bar{q}$ interaction in the vector channel has become
very weak, signaling the onset of deconfinement. 
This also indicates the quark-hadron duality~\cite{shuryak1999,rapp1999} 
near the transition point.


It is well known that 
the $p_T$ spectra of photons and lepton pairs are sensitive 
to the values of initial temperature $T_i$, $v_0$,
$T_f$ and EOS. The $T_c$ dependence of the $p_T$ distribution  
is found to be negligibly small~\cite{anrms}. As we have mentioned before 
though these parameters can be constrained 
from the measured multiplicity and freeze-out spectra 
there remains  still
some room to vary these quantities 
in order to be able to 
describe the experimental data. 

The $p_T$ dependence of the ratio, $R_{em}$ for SPS, RHIC and LHC energies are 
shown in Figs.~\ref{fig2},~\ref{fig3} and ~\ref{fig4} respectively. 
It is observed that at a given $p_T$, the ratio decreases with $M$,
reaches a minimum around $\rho$-peak and increases beyond the $\rho$-peak.
This trend is valid for all the cases, {\it i.e.} SPS, RHIC and LHC as
expected because at a given $p_T$ the $R_{em}$ is actually the inverse of the 
invariant mass  distribution of lepton pairs (the denominator {\it i.e.}
the photon spectra is
same for all the mass windows).  
It is observed from Figs.~\ref{fig2},\,\ref{fig3} and \ref{fig4}
that the ratio, $R_{em}$ decreases
with $T_i$ for given $p_T$ for $M$ below the $\rho$-peak and the opposite
behaviour is observed above the $\rho$-peak. The slope of the ratio at
low $p_T$ also indicates  substantial change with increasing $M$,
the slope is minimum at the $\rho$-peak. Therefore, the minimum of the
slope may be used to locate the effective mass of the vector meson 
in medium.  

It is clear from the results displayed in Figs.~\ref{fig2}
~\ref{fig3} and ~\ref{fig4}
that the quantity, $R_{em}$, reaches a plateau beyond $p_T=1.5$ GeV
for all the three cases {\it i.e.} for SPS, RHIC and LHC.
It may be noted here that the degree of flatness increases
from SPS to RHIC and LHC.  
As mentioned before for all the three cases, except
$T_i$  all other quantities {\it e.g.} $T_c$, $v_0$ and EOS
are same, so the difference in the value of $R_{em}$ in the plateau
region originates due to different  values of initial temperature, 
indicating  this can be a measure of $T_i$. 

The following analysis will be useful
to understand the origin of the plateau at high $p_T$ region.
The strong three momentum dependence in the dilepton and 
photon emission rates (Eqs.~\ref{virtual} and ~\ref{real} respectively)
originates from the thermal factor, $f_{BE}(E,T)$. 
For a static system the energy, $E$ can be written as 
$E=M_Tcoshy$, where $M_T=\sqrt{p_T^2+M^2}$ $y=tanh^{-1}{p_z/E}$. 
At high $p_T (>>M)$, $M_T\approx p_T$, the exponential
momentum dependence become same for real photon $(M^2=0)$ and dilepton 
$(M^2\neq 0)$ spectra  and
hence plateau is expected in the static ratio for large $p_T$ for 
all the $M$ values. 

We recall that for an expanding system out of the two
kinematic variables describing the dilepton spectra,  $p_T$ is affected
by expansion  but $M$ remains unchanged. The range of $M$ under present
study is $0.3<M\mathrm{(GeV)}<1.3$. 
The energy, $E$ appearing in both the photon and dilepton 
emission rates should be replaced by $u^\mu p_\mu$
for a system expanding with space-time dependent four velocity $u^\mu$.
Under the assumption of cylindrical symmetry and longitudinal boost
invariance $u^\mu$ can be written as
\be
u^\mu=\gamma_r(t/\tau,v_rcos\phi,v_rsin\phi,z/\tau)
\ee
where $\tau=\sqrt{t^2-z^2}$ $t=\tau cosh\eta$, $z=\tau sinh\eta$,
$v_r(\tau,r)$ is the radial velocity, 
$\gamma_r(\tau,r)=(1-v_r(\tau,r))^{-1/2}$.
The four momentum,
$p^\mu=(M_Tcoshy,p_T,0,M_Tsinhy)$ where
$p_L=m_Tsinhy$. Therefore, for dilepton 
\be
u^\mu p_\mu=\gamma_r(M_Tcosh(y-\eta)-v_rp_Tcos\phi)
\label{flowe}
\ee
for photon the factor 
$u^\mu p_\mu$ can be obtained by replacing $M_T$ in Eq.~\ref{flowe}
by $p_T$.
The $p_T$ dependence of the photon
and dilepton spectra originating from an expanding 
system is predominantly determined by the thermal factor
$f_{BE}$.  Therefore, we discuss following three scenarios.
(i)At high $p_T (>>M)$, $M_T\approx p_T$, the exponential
momentum dependence become same for real photon and dilepton spectra, 
hence for large $p_T$ a plateau is obtained in the ratio, $R_{em}$
(Figs~\ref{fig2}-\ref{fig4}). 
In other words, the effect of radial flow on the photon and
dilepton is similar at high $p_T$ region. 
(ii)If the large $M$ pairs originate from early time 
(when the flow is small) 
the ratio, $R_{em}$ which includes space-time dynamics
will be close to the static case and hence will show 
plateau.
(iii) However, at late time when the radial flow is large and
$M$ is comparable to or larger than $p_T$ the effect of flow
on dilepton will be larger (receives larger radial kick due to
non-zero $M$) than the photon and hence
the plateau may disappear. Therefore, the disappearance of
plateau structure in $R_{em}$ in moderate or high $M$ region
will indicate the presence large radial flow. This can be understood
from the results shown in Figs ~\ref{fig5} and ~\ref{fig6}.
 
In Fig.~\ref{fig5} the ratio has been displayed 
only for quark matter. Here the flow is
expected to be lower within the present framework of present study.
A plateau is observed for all the $M$ windows.
It is observed that for high $M$ ($\sim 1.2$ GeV) and low $M$ ($\sim 0.3$
GeV) the ratio for QM is close to the total for LHC energy (not shown
separately). 

In~Fig.\ref{fig6} the ratios has been displayed 
for hadronic matter only. Here the flow is expected to be
very large. Within the ambit of the present modeling
the  contribution from the hadronic matter is overwhelmingly
large in the $M$ region, $0.7< M < 0.8$ GeV. Therefore, this
region will have large effects from the radial flow and hence
it may destroy the plateau. This is clearly seen in Fig.~\ref{fig6}
for the curve corresponding to $0.7< M < 0.8$ GeV. 

To demonstrate the effect of flow on the plateau we use an 
initial velocity profile (which gives rise to stronger radial flow
than Eq.~\ref{eq8}) of the form $v_r(\tau_i,r)=v_0^\prime\frac{r}{R_A}$
with an {\it unrealistically} large value of $v_0^\prime\sim 0.5$. These
inputs are used only for results shown in Fig.\ref{figflow}, which
clearly indicates the disappearance of plateau. Variation of $R_{em}$
with $p_T$ corresponding to hadronic phase is steeper than the total
because of larger radial flow in the late stage of the evolution. 

Now we demonstrate the effect of other parameters on $R_{em}$.
The value of $T_c$
has large uncertainties. Therefore, we show the sensitivity
of the results on $T_c$ in Fig.~\ref{fig7} for two invariant mass windows.
The results are insensitive to $T_c$. 

The effect of the EOS
on $R_{em}$ is demonstrated in Fig.~\ref{fig8}, by varying $\Gamma$ in 
Eq.~\ref{eos}. 
It is observed that the effect of EOS on $R_{em}$ for both
the mass windows are small. Similar to the effect of $T_c$,
the larger mass window ($1.2\leq M$(GeV)$<1.3$) is less
affected by the change in EOS. This is because the effect
radial flow (and other hydrodynamic effects) are less at
early times from where higher mass lepton pairs originate.
Replacement of lattice
QCD EOS for QGP phase by bag model shows negligible effects
on $R_{em}$. 

In Fig.~\ref{figti} the dependence of $R_{em} (p_T=2.5\mathrm{GeV})$ 
is depicted as a function of $T_i$ for $1.2\leq M$(GeV)$<1.3$. 
This mass window is selected  because the contributions 
from the hot quark matter phase  dominates this region
and the effects of $T_c$, EOS etc are least here.
$p_T=2.5$ GeV is taken because $R_{em}$ achieved 
a complete plateau at this value of transverse momentum. 
The change in $R_{em}$ from SPS to RHIC is about $40\%$ and
from RHIC to LHC this is about $20\%$.  A simultaneous measurements
of photons and dileptons with required accuracy,
will be useful to disentangle the effects of flow and true average
temperature in a space-time evolving system formed in heavy ion
collisions at ultra-relativistic energies.

\begin{figure}
\begin{center}
\includegraphics[scale=0.45]{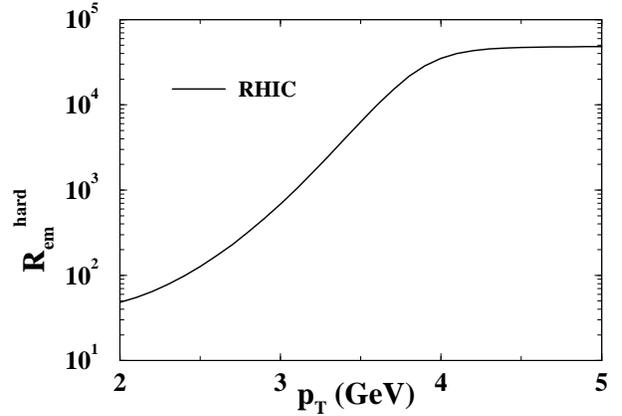}
\caption{The variation $R_{em}$ for hard photons to dileptons 
ratio as a function of $p_T$ for  $\sqrt{s_{NN}}=200$ GeV and
invariant mass window, $M=0.2-0.3$ GeV.
}
\label{fig9}
\end{center}
\end{figure}

We have evaluated $R_{em}^{pQCD}$, the ratio 
$(d^2N_\gamma/d^2p_Tdy)_{y=0}/(d^2N_{\gamma^\ast}/d^2p_Tdy)_{y=0}$
for hard processes using pQCD (Fig.~\ref{fig9}).
The hard photon contributions has been constrained to
reproduce the PHENIX data~\cite{ppPHENIX} for $pp$ 
collisions at $\sqrt{s_{NN}}=200$ GeV.
We consider 
$q \bar{q} \rightarrow \gamma ^{*}\rightarrow l^{+}l^{-}$,
$q$$\bar{q}$$\rightarrow$$g$$\gamma^{*}$ and 
$q$$g$($\bar{q}$)$\rightarrow$$q$$\bar{q}$$\gamma^{*}$
for the lepton pair production. The $M$ integration 
of lepton pair spectra (Eq.~\ref{eq3}) 
is done over the range $0.2 \leq\,M$(GeV)$ \,\leq \,0.3$.
We observe that $R_{em}^{pQCD}$ increases for $p_T$ up to $\sim 3 $ GeV, above
which it reaches a plateau.
Therefore, for  $p_T\sim 1-3$ GeV, $R_{em}$ for the
thermal and pQCD processes show different kind of behaviour.  
The plateau
arises from the fact that at large $p_T$ both photon
and dilepton show power law behaviour~\cite{Ellis,Field}.
In the low $p_T$ domain  lepton pairs (photon) from pQCD processes indicate 
a Gaussian type~\cite{Ellis} (power law) variation resulting in the
increase of $R_{em}$ with $p_T$.  

\section{Summary and Conclusions}
We have studied the variation of $R_{em}$, the ratio  of the 
transverse momentum spectra of photons and
dileptons and argued that measurement
of this quantity will be very useful to determine the
value of the initial temperature of the system formed
after heavy ion collisions. We have observed that 
$R_{em}$ reaches a plateau beyond $p_T= 1.5$ GeV. The
value of $R_{em}$ in the plateau region depends on 
$T_i$.  However, the  effects of flow, EOS and the dependence on 
the values of  $T_c$, $v_0$ and other model
dependences get canceled away in the ratio, $R_{em}$. 
For $M$ above and below the $\rho$ peak and $p_T\geq 2$ GeV
the contributions from quark matter dominates, therefore
these regions could be chosen to estimate the initial temperature
of the system formed after the collisions.

It is well known that $T_{eff}$, the inverse slope 
(see {\it e.g.}~\cite{na60prl2})
extracted from the  $p_T$ spectra of 
EM  radiation contains the effect of temperature as well as flow. 
We have seen that when the flow is less (in the  initial stage of 
the evolution) the
ratio, $R_{em}$  shows a plateau for large $p_T (>>M)$, the height of
the plateau in this region will give a good measure of the average
temperature. However, a large flow can destroy the plateau and
hence the deviation from the flatness of the $R_{em}$ versus
$p_T$ curve may be used as a measure of flow. 
So a careful selection  of $M$ and $p_T$ regions will
be very helpful to disentangle the effect of true average
temperature and the flow (see also~\cite{renk}).  In~\cite{renk}
it was shown that the effects of flow on high $M (>1.5$ GeV) could
be quite large and in such cases the plateau in $R_{em}$ may 
disappear. However, in the present work we confine in
the  range $0.3<M){\mathrm{GeV}})<1.2$.


EM radiations originating from the interactions
between thermal and non-thermal (high energy) 
partons~\cite{npa1997,jetconv,LB}
has been neglected in the present work. It is expected that
the EM radiation from these processes and also from the
pre-equilibrium stage will not affect $R_{em}$.

We have studied the effects of chemical off-equilibrium of mesons on the
photon and dilepton production rates. This is
implemented by appropriately introducing non-zero pionic chemical 
potential, $\mu_\pi$ ($\mu_\rho=2\mu_\pi, \mu_\omega=3\mu_\pi$)
in the thermal factors~\cite{HR} appearing both in   
photon and dilepton emission rates. We observed that the plateau structures
in $R_{em}$ do not change for RHIC and LHC, but for SPS it has little effect.

The change in hadronic spectral function at non-zero temperature 
and density is a field of high contemporary research interest as 
this is connected with the restoration of chiral symmetry in QCD.
From the QGP diagnostics  point of view the  
background contributions (photons and dileptons from thermalized hadrons) 
are affected due to medium effects on hadrons. Therefore, some comments
on this issue are in order here. 

We have checked that the $p_T$ spectra of both
photons and  dileptons
are  sensitive to the  pole shift of hadronic spectral function,
as the reduction of hadronic masses~\cite{BR} in
a thermal bath increases their abundances
and hence the rate of
emission gets enhanced~\cite{alam2,we1,we2,ja1,ja2}. 
The invariant mass distribution of lepton pairs are 
sensitive to both the pole shift and broadening~\cite{rw,ja2,sah,li,rapp}.
But the $p_T$ spectra of the EM radiation is insensitive 
to the broadening of the spectral function provided the 
integration over the $M$  is performed 
over the entire region. This is because broadening
does not change the density of vector mesons 
significantly (see also~\cite{ja2}).
However, the number density of vector mesons depends
on the nature (shape) of the spectral function within the
integration limit.  Therefore,
the $p_T$ spectra may change due to broadening 
when the integration over $M$ is done in a limited $M$ domain.
We have checked that doubling the $\rho$ width ($\sim 2\times 150$ MeV) 
changes $R_{em}$ by $10\%$.  
It is important to note that the change in
mass and widths can not be arbitrary it should obey certain constraints
as discussed in~\cite{leupold}.
Therefore, simultaneous measurements of $p_T$ spectra and invariant
mass distribution of real and virtual photons
could be very useful to understand the nature 
of medium effects on hadrons~\cite{ja2}.

{\bf Acknowledgment:} We are grateful to  Ludmila Levkova for providing the 
lattice QCD results for Equation of state.

\end{document}